# Graph Computing based Distributed Fast Decoupled Power Flow Analysis

Chen Yuan[a], Yi Lu[b], Wei Feng[c], Guangyi Liu[a], Renchang Dai[a], Yachen Tang[a], Zhiwei Wang[a]
[a] GEIRI North America, San Jose, CA, USA
[b] State Grid Sichuan Electric Power Company, Chengdu, Sichuan, China
[c] The University of Tennessee, Knoxville, TN, USA

*Abstract*—Power flow analysis plays a fundamental and critical role in the energy management system (EMS). It is required to well accommodate large and complex power system. To achieve a high performance and accurate power flow analysis, a graph computing based distributed power flow analysis approach is proposed in this paper. Firstly, a power system network is divided into multiple areas. Slack buses are selected for each area and, at each SCADA sampling period, the inter-area transmission line power flows are equivalently allocated as extra load injections to corresponding buses. Then, the system network is converted into multiple independent areas. In this way, the power flow analysis could be conducted in parallel for each area and the solved system states could be guaranteed without compromise of accuracy. Besides, for each area, graph computing based fast decoupled power flow (FDPF) is employed to quickly analyze system states. IEEE 118-bus system and MP 10790-bus system are employed to verify the results accuracy and present the promising computation performance of the proposed approach.

## I. Introduction

Power systems are becoming more interconnected and complicated networks over the last decades. Advanced power electronic devices enable high penetrations of renewable energy resources, distributed generation, energy storage system, responsive loads, and electric vehicles at both transmission and distribution levels [1]–[4]. Besides, the flexible alternating current transmission system (FACTS), like static VAR compensator and static synchronous compensator (STATCOM), are also adapted to actively control power flow with the use of power electronics interfaces. Furthermore, hybrid AC/DC systems are also rapidly developed in modern power grids [5]. Their integrations inevitably and dramatically increase the power grid's complexity and bring in more frequent fluctuations and uncertainties, challenging the computation speed of power flow applications in commercial EMS. Power flow analysis, as a fundamental and critical function in EMS, is required to accommodate large-scale and highly complicated modern power systems as fast as possible, even with the capability of situation awareness and potential contingency prediction [6], [7].

In the last decades, power system society has endeavored to improve power flow analysis from the data structure to algorithms to speed up its computation performance, especially for the large-scale system. Meanwhile, with power system modernization, an advanced system architecture and a fast computation approach are indeed needed to maximally guarantee power system operation robustness, reliability, security, and resilience. Parallel computing is one promising method to improve computation efficiency by taking advantages of advanced computation technique, rich storage space and parallel capability of processing units. However, the state of art of power flow analysis in commercial EMS does not effectively make use of the parallel computation capability, since the relational database and computation algorithm used for existing power flow analysis and EMS were not specifically designed for parallel computing. With the evolution of database, data structure and hardware configuration, the external conditions of parallel power flow analysis become mature. Reference [8] used distributed computation technology to implement parallel power flow computation. Besides, GPU based parallel computing was introduced and applied to power flow calculation [9], [10].

On the other hand, with the fast development of computing technology and graph theory based applications, graph-based high-performance computation, i.e. *graph computing*, is a feasible option for high-performance computing. It has been developed to deal with distributed storage and parallel computing in big data analysis, and applicable to solve complicated scenarios with iterations [11]. Graph-based power flow calculation method was presented to demonstrate its advantages in power flow analysis [12]. In addition, previous works also investigated the feasibility and the high parallel computation performance of graph-based power system EMS applications, like state estimation, power flow analysis, "N-1" contingency analysis, and security constrained economic dispatch [13]–[16]. However, they took the entire system as an entity and do the computation in a centralized way.

In this paper, a graph computing based distributed FDPF approach is proposed to speed up the computation performance without the compromise of results accuracy. At first, a power system network is divided into multiple areas. Slack buses are selected for each area and, at each SCADA sampling period,

This work is supported by the State Grid Corporation technology project 5455HJ180020.

slack bus angle differences are recorded from EMS state estimator. Furthermore, the power flows of the inter-area transmission line are equivalently allocated as extra load injections to corresponding buses. Then, the system is transformed to multiple independent areas. In this way, the power flow analysis could be conducted in parallel for each area and the solved system states could be guaranteed without compromise of accuracy. In addition, graph computing based FDPF is employed to quickly calculate system states for each area.

This paper is organized as follows. In Section II, the graph computing and system partitioning are briefly introduced. Then, the graph computing based distributed FDPF method is elaborated in Section III. Case study is conducted in Section IV to verify the results accuracy of the proposed approach and demonstrate its significant computation efficiency.

## II. Graph Computing and System Partitioning

### A. Graph Database and Graph Computing

Graph is a data structure modeling pairwise relations between objects in a network. In mathematics, a graph is represented as $G = (V, E)$, in which $V$ indicates a set of vertices, representing objects, and the set of edges is represented as $E$, expressing how these vertices relate to each other. Each edge is denoted by $e = (i, j)$ in $E$, where $i$ and $j$ in $V$ are referred as head and tail of the edge $e$, respectively [17].

Previous works have explored the usage and feasibility of the graph database to naturally represent power systems and apply graph computing to energy management systems in power grids [18]. In this subsection, two main parts of graph computing, i.e. node-based parallel computing and hierarchical parallel computing, are presented as follows.

*1) Node-based Parallel Computing*: In graph computing, node-based parallel computing represents that the computation at each node is independent and can be conducted in parallel. In Fig. 1, its upper half depicts the mapping relation between graph-based computation and matrix computation. It clearly shows that, in a graph, the counterparts of the connections between nodes are non-zero off-diagonal elements in the coefficient matrix, $A$. Zero off-diagonal elements in the coefficient matrix indicate that no direct connections between the nodes exist in the graph. The bottom half of Fig. 1 demonstrates the node-based parallel computing strategy. Taking the admittance matrix in Fig. 1 as an example, the off-diagonal element is locally calculated based on the impedance attributes of the corresponding edge, and each diagonal element is independently obtained only with the processing of impedance attributes at the corresponding node and its connected edges. Therefore, the whole admittance matrix is developed with one-step graph operation and the value of each element is calculated independently and in parallel. Other examples of node-based parallel computation in power systems are active/reactive power injection calculation, node variables mismatch update, active/reactive power flow calculation, etc.

*2) Hierarchical Parallel Computing:* Hierarchical parallel computing performs computation for nodes at the same level

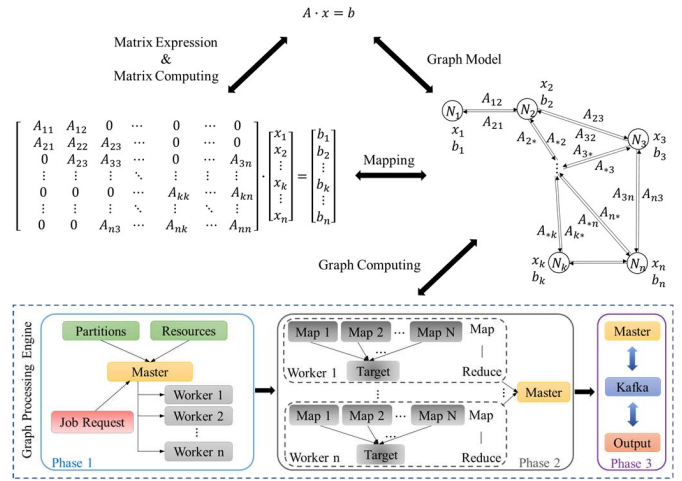

Figure 1. Mapping between graph computation and matrix computation

in parallel. The level next to it is performed after. An example of hierarchical parallel computing application is matrix factorization. Cholesky elimination algorithm is employed to conduct matrix factorization. Three steps are involved: 1) determining fill-ins, 2) forming an elimination tree, and 3) partitioning the elimination tree for hierarchical parallel computing.

### B. System Partitioning for Distributed Graph Computing

In the previous subsection, the concept of graph computing is presented. But, as it depicts, it takes the entire system as an entity and conducts the processing together. All the nodes do the local computing first, then communicate and convey information with others, and at last wait at the barrier for synchronization.

To further speed up the power flow analysis performance, system partitioning is used for distributed graph computing. Since the paper mainly focuses on implementing distributed power flow analysis with developed system partitioning [9], [19], the exact partitioning approach is out of scope. After the system partitioning, the overall system is decomposed into multiple non-overlapping areas, so that each area could be processed independently and in parallel using graph computing technique. In other words, each area is processed independently, and, within each area, graph computing is applied to implement parallel processing. So, the burden of local computation, communication, and synchronization for each area is much reduced, as shown in Fig. 1.

## III. Graph Computing based Distributed Fast Decoupled Power Flow

In this section, the approach of distributed graph computing will be applied to power flow analysis and the distributed FDPF analysis using graph computing is illustrated.

### A. Graph Computing based Fast Decoupled Power Flow

Fast decoupled power flow is a widely used approach for power flow analysis. Its problem formulation is shown below.

$$\begin{cases} B' \cdot \Delta\theta = \Delta P/|V| \\ B'' \cdot \Delta|V| = \Delta Q/|V| \end{cases} \quad (1)$$

where, $B'$ and $B''$ are approximated Jacobian matrices, assuming $cos\theta_{ij} \approx 1, R_{ij} \ll X_{ij}$, $|V|$ is a vector of bus voltage magnitudes, $\Delta\theta$ and $\Delta|V|$ are vectors of bus voltage phase angle and magnitude incremental. Similarly, $\Delta P/|V|$ and $\Delta Q/|V|$ are vectors of active and reactive power injection incremental divided by corresponding bus voltage magnitude.

The flowchart of graph computing based FDPF is presented in Fig. 2. Portions of the approach are highlighted with blue outlines, indicating that they are implemented with node-based parallel computing. Meanwhile, the method of hierarchical parallel computing is applied to the rest parts of the approach, which are surrounded by green squares in Fig. 2. The following will further explain how to implement the FDPF analysis with graph computing technique.

*1) Node-based parallel computing – formulating power flow equation, calculating branch flow and updating system states:* Since $B'$ and $B''$ are approximated Jacobian matrices, non-zero off-diagonal elements indicate existing connections and diagonal elements correspond to nodes in the system network. Taking a further look, it is not difficult to find that each row vector is related to a corresponding node in the system. In each row vector, non-zero off-diagonal elements are linked edges from the corresponding node and the diagonal element represent the node itself. Thus, all the elements in $B'$ and $B''$ can be acquired locally and independently from others, indicating the feasible application of node-based parallel computing. Besides, in equation (1), the right-hand-side vectors, $\Delta P/|V|$ and $\Delta Q/|V|$, could also be updated using node-based parallel computing. This is because $\Delta P$, $\Delta Q$ and $|V|$ are attributes of each node in the system graph and the computation could be definitely done locally and in parallel. Except for problem formulation, branch power flow calculation and system states update are also locally conducted with node-based parallel computing. This is because branch power flow only needs information from the corresponding node, its neighbors and linked edges, and the update of system states, i.e. bus voltage magnitude and phase angle, are directly implemented at each local node in the system graph.

*2) Hierarchical parallel computing – solving power flow:* After the formulation of the power flow problem, an efficient solver is needed to quickly find the accurate solution. As shown in Fig. 2, green squares are the core members of the FDPF solver and they are implemented with hierarchical parallel computing to achieve high computation performance. Nodes at the same level are analyzed and calculated in parallel and the level next to it is performed after.

*B. Graph Computing based Distributed Fast Decoupled Power Flow Approach*

To further improve the performance of power flow analysis, system portioning into multiple areas will be described in this subsection, and then followed by the proposed distributed power flow method using the skill of graph computing.

In the conventional distributed approach, system network is first divided into multiple areas based on the geological information and topology structure [19]. Then, each area performs its own power flow and coordinating with others to find the solution by exchanging information at the border. In this paper, the divided multiple areas are equivalently considered as partitioned areas isolated from others. Slack buses are selected for each area, and, at each SCADA sampling period, the power flows of inter-area branches are equivalently represented by extra active power and reactive power injections to corresponding buses. Then the power flow in each area is calculated independently. In other words, no more information exchange is needed at borders of different areas during the execution of power flow analysis, since the boundary information is preprocessed and exchanged before power flow calculation. On the other hand, with the use of state estimator, the inter-area branch power flow and voltage phase angles at slack buses are determined and then fixed within each SCADA sampling period. They will be periodically updated when new SCADA signal comes and state estimation executes. As shown in Fig. 3, IEEE 14-bus system is divided into four areas [19]. There are 7 inter-area branches. Taking branch 4-5 as an example, it is interconnecting area 1 and area 2, and also connecting bus 4 and bus 5. During the process of system network partitioning, branch 4-5 is removed and its power flows, $P_{4-5}$, $Q_{4-5}$, $P_{5-4}$ and $Q_{5-4}$ ($P_{4-5}+P_{5-4} \approx 0$ and $Q_{4-5}+Q_{5-4} \approx 0$), are equivalently replaced by extra power injections at bus 4 and bus 5, which are $P_{4-5}, Q_{4-5}, P_{5-4}$ and $Q_{5-4}$, respectively. Similarly, the rest of inter-area branches are equivalently replaced by extra load injections. After the inter-area branches removed and equivalent extra load injected to corresponding buses, the IEEE 14-bus system is partitioned into four isolated areas, as shown in Fig. 3. The buses with dashed squares were once connected with inter-area branches. So, they have equivalent extra load injections after system partitioning. Since IEEE 14-bus system is a very small system and its admittance matrix is also kind of dense, the ratio of impacted buses after system partitioning is high. But, for larger systems, like IEEE 118-bus system and MP 10790 system, buses are not tightly connected with others and we will see the ratio of impacted buses is very low in Section IV.

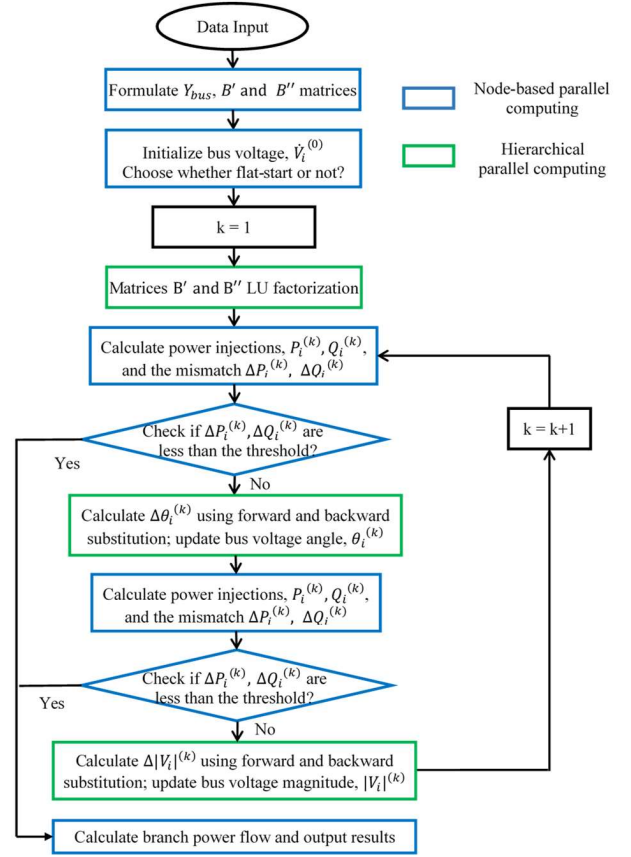

Figure 2. Flowchart of graph computing based fast decoupled power flow

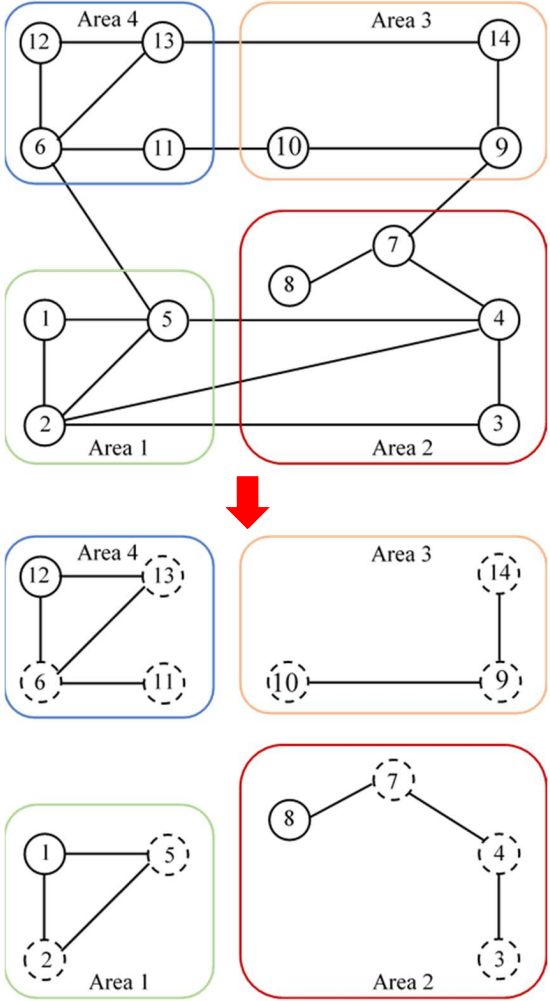

Figure 3. IEEE 14-bus system partitioning with equivalent extra load injections

After the system network partitioning, the distributed graph computing based FDPF analysis can be applied to all areas simultaneously.

## IV. PERFORMANCE EVALUATION AND DISCUSSION

### A. Testing Environment and Testing Cases

To verify the results accuracy and demonstrate the high computation performance of the proposed approach, the testing is conducted in a Linux server with the installation of a graph database platform. The detailed configurations of the testing environment are listed below in Table I.

TABLE I. TESTING ENVIRONMENT

| Hardware Environment | |
|---|---|
| CPU | 2 CPUs × 6 Cores × 2 Threads @ 2.10 GHz |
| Memory | 64 GB |
| Software Environment | |
| Operating System | CentOS 6.8 |
| Graph Database | TigerGraph v0.8.1 |

In the following subsections, the standard IEEE 118-bus system, which is a simple approximation of the American Electric Power system in the United States midwest area, is used to first verify the accuracy of the proposed approach, and MP 10790-bus system [9], which is a system with four interconnected European systems, is employed to demonstrate the promising computation performance.

### B. Approach Verification

Table II provides the comparison results with MatPower, using IEEE 118-bus system. The system partitioning of the IEEE 118-bus system is displayed in Fig. 4 [19]. The ratio of impacted buses by equivalent extra load injections is only 11%.

TABLE II. ACCURACY VERIFICATION USING IEEE 118-BUS SYSTEM

| Method | Number of Threads | Computation Time (ms) | Max Bus Voltage Difference from MatPower | |
|---|---|---|---|---|
| | | | Phase Angle (degree) | Magnitude (per unit) |
| Proposed Method | 1 | 2.23 | 0.0068 | 0.00054 |
| MatPower FDPF | 1 | 48.30 | — | — |

In Table II, both tests are initialized with non-flat start, and the convergence criteria for the mismatch of bus active and reactive power injections are set at 0.001 p.u. Regarding the results accuracy, the maximum bus voltage phase angle difference is 0.0068 degree and the maximum magnitude difference is 0.00054 in per unit. Both are negligible. The computation time of the proposed method is less than 5% of the time spent in MatPower. On the other hand, for this case, only 1 thread is used since it is a small system. With more running threads, the time cost is increasing because of the communication time and overhead cost outweighs the time saving of parallel computing.

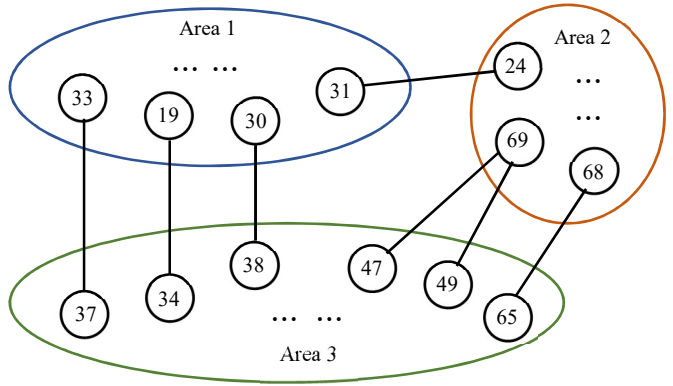

Figure 4. IEEE 118-bus system partitioning

*C. High Computation Performance*

After accuracy verification, the testing to demonstrate the proposed method's promising computation efficiency with multiple threads is presented in this subsection. As mentioned before, it is difficult to show a better parallel computation performance with a small system. Therefore, MP 10790-bus system is used to test the parallel computing performance of the proposed method with non-flat start. As displayed in Fig. 5, the system is partitioned into 4 areas [9]. The ratio of impacted buses by equivalent extra load injections is less than 0.06%.

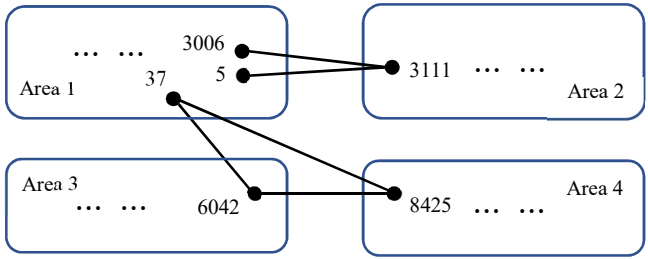

Figure 5. MP 10790-bus system partitioning

The testing performance is presented in Table III. Not only the parallelism testing of the proposed method is conducted, but also the computation performance of the graph computing based FDPF method without system partition [12] is presented. From Table III, it is not hard to obtain the conclusion that (a) with the increase of running threads, computation performance is greatly improved, and (b) the performance of the proposed method is much better than the graph computing based FDPF. Regarding the conclusion (a), it demonstrates the promising computation capability of graph computing based approaches. In the conclusion (b), it emphasized the improved efficiency with smaller matrix operation and less computation effort for each area power flow analysis. Besides, the simultaneous power flow analysis for multiple areas largely reduced the time cost.

TABLE III. PARALLEL COMPUTATION PERFORMANCE TESTING WITH MP 10790-BUS SYSTEM

| Method | Number of Iterations | Computation Performance Under Different Number of Running Threads (ms) | | | |
|---|---|---|---|---|---|
| | | 1 | 2 | 4 | 8 |
| Graph Computing based FDPF | 4 | 138.06 | 110.49 | 94.63 | 88.82 |
| Proposed Method | Area 1: 4 Area 2: 4 Area 3: 3 Area 4: 3 | 94.87 | 55.92 | 35.02 | 31.42 |

## V. CONCLUSION

In this paper, a graph computing based distributed fast decoupled power flow approach was presented to further speed up the computation performance without the compromise of results accuracy. With system network partitioning and inter-area branch power flow being substituted by equivalent extra load injections, a system could be divided into multiple isolated areas and distributed power flow analysis could be applied to each area with graph computing technique. IEEE 118-bus system and MP 10790-bus system are employed to verify the results accuracy and display the significant computation efficiency of the proposed method.